\documentstyle[multicol,eqsecnum,aps,prd,epsf]{revtex}
\def\dA{d_{\rm A}}
\def\rH{r_{\rm H}}
\def\ms{m_{\rm s}}
\def\NH{N_\protect{\rm H\protect}}
\def\NT{N_\protect{\rm T\protect}}
\def\Hi{H_\protect{\rm i\protect}}
\def\Hsi{H_\protect{\rm shell\protect}}
\begin{document}
 \makeatletter
 \newdimen\ex@
 \ex@.2326ex
 \def\dddot#1{{\mathop{#1}\limits^{\vbox to-1.4\ex@{\kern-\tw@\ex@
  \hbox{\tenrm...}\vss}}}}
 \makeatother
\thispagestyle{empty}
{\baselineskip0pt
\leftline{\large\baselineskip16pt\sl\vbox to0pt{\hbox{Department of Physics} 
               \hbox{Kyoto University}\vss}}
\rightline{\large\baselineskip16pt\rm\vbox to20pt{\hbox{KUNS-1514} 
\vss}}%
}
\vskip1cm
\begin{center}
{\large{\bf Distance--redshift relation in 
an Isotropic Inhomogeneous Universe}}\\
{\large --- Spherically Symmetric Dust--shell Universe ---} \\
\end{center}
\begin{center}
{\large Norimasa Sugiura\footnote{E-mail:sugiura@tap.scphys.kyoto-u.ac.jp}, 
Ken-ichi Nakao\footnote{nakao@tap.scphys.kyoto-u.ac.jp}
and Tomohiro Harada\footnote{harada@tap.scphys.kyoto-u.ac.jp}}\\
{\em Department of Physics,~Kyoto University, 
 Sakyo-ku, Kyoto 606-8502, Japan}
\end{center}
\begin{abstract}
The relation between angular diameter distance and redshift 
($\dA$--$z$ relation) 
in a spherically symmetric dust--shell universe is studied.
This model has large inhomogeneities
of matter distribution on small scales.
We have discovered that the relation agrees with that
of an appropriate Friedmann--Lema\^{\i}tre(FL) model
if we set a ``homogeneous'' expansion law and a ``homogeneous''
averaged density field. 
This will support the averaging hypothesis that a universe
looks like a FL model in spite of small-scale fluctuations
of density field,
if its averaged density field is homogeneous on large scales.
\end{abstract}


\section{Introduction}

The standard big bang model is based on the assumption of 
homogeneous and isotropic distribution of matter and 
radiation. This assumption then leads to the Robertson-Walker 
(RW) space-time geometry and the Friedmann-Lema\^{\i}tre (FL) 
universe model\footnote{We use the term ``Robertson-Walker
 space-time'' when we focus on geometrical aspects of a homogeneous
 and isotropic model, and ``Friedmann-Lema\^{\i}tre model'' when we
discuss its dynamics and observable quantities.}
 through the Einstein equations. 
This model has succeeded in explaining various important observational 
facts: the Hubble's expansion law, the contents of light elements and 
the isotropy of cosmic microwave background radiation 
(CMBR)\cite{rf:WEINBERG}. 

The CMBR conversely gives a strong 
observational basis for the assumption of homogeneity and 
isotropy of our universe by its highly isotropic 
distribution together with the Copernican 
principle.
Indeed, the deviation 
of our universe from a homogeneous and isotropic space 
is as small as $10^{-5}$\cite{rf:COBE}
at the stage of decoupling.
Thus our universe is well approximated by 
a FL model before this stage. 
On the other hand, the present universe is highly inhomogeneous 
on small scales; the density contrast is
of the order of $10^{30}$ in the solar system,
$10^{5}$ on galactic scales,
and of the order of unity even on the scale of superclusters. 
We have to go beyond FL models 
in considering such systems.

For a long time we have regarded that
a FL model is a large-scale ``average'' of a
locally inhomogeneous universe (averaging hypothesis). 
Even though the observational data are consistent with the 
picture that our universe is described well by 
a RW metric, we are not sure 
how to derive the background FL model from the inhomogeneous universe
by any averaging procedure,
or how the non-linear inhomogeneities on small scales 
affect large-scale behavior of the universe\cite{rf:ELLIS}.
Although one can derive a background 
FL model from observations of the nearby galaxies
with any rule of averaging one likes,
it is uncertain
whether this background FL model agrees with the FL model
whose cosmological parameters was defined 
at the stage of decoupling.
The discrepancies might appear, for example, in the 
estimate of the density of baryonic matter, 
the density parameter, the age of our universe, and so on.
These still remain a non-trivial question 
to which we have to give a clear answer.

A number of approaches have been made to study how the small-scale  
inhomogeneities affect the global dynamics when averaged 
on larger scales\cite{rf:FUTAMASE,rf:BE,rf:RSKB,rf:CP}. 
The first work which explicitly showed the existence of 
such an effect was performed by Futamase\cite{rf:FUTAMASE}. 
Assuming small deviation of the space-time geometry from 
the RW one, Futamase constructed an elegant formalism 
by the use of the Post-Newtonian expansion and 
Isaacson's prescription to take into account the back reaction 
of the small-scale inhomogeneities on the global cosmic 
expansion. 
%
After his works, Buchert and Ehlers\cite{rf:BE} studied
this back reaction problem in the framework of the Newtonian
cosmology in  which the corresponding background FL model is 
uniquely determined through the spatial averaging of physical quantities 
without any uncontrolled approximation.
They showed that inhomogeneities do not influence the 
overall expansion in spatially compact models 
(the topology of its spatial section is $T^{3}$) 
if they are averaged over 
the whole space. In other cases, the significance of 
inhomogeneities may depend on the cosmology one adopts. 
Another interesting approach has been made by Carfora and 
Piotrkowska\cite{rf:CP}, in which 3-dimensional geometry
is deformed according to the Ricci-Hamiltonian flow
which they say would be equivalent to changing the scale of
averaging. They derive a homogeneous geometry
which corresponds to a large-scale average of an
inhomogeneous universe, and discussed the 
effect of inhomogeneities on the cosmological parameters.

In spite of these works, the effect of inhomogeneities
on the cosmological parameters remains unclear;
there are even apparent discrepancies in their statements.
The reason seems to lie in the different definition of
back reaction of inhomogeneities\cite{rf:KASAI};
its definition is ambiguous since
they do not treat observable quantities in these works. 
Thus, in order to understand clearly the effects of inhomogeneities, 
it is necessary to relate them with physical quantities 
which we can give an unambiguous definition.

Bildhauer\cite{rf:BILD} showed that the global cosmic expansion 
rate is not isotropic if the back reaction of
small-scale inhomogeneities are taken into account
and then investigated its observational effects on the CMBR. 
After this work, Bildhauer and Futamase\cite{rf:BF} 
discussed the possibility that the observed 
dipole anisotropy in the CMBR comes from this effect. 
These works are significant in the sense that the back reactions 
on the observable quantities were discussed,
but they did not consider the cases where we try to 
determine the cosmological parameters by observing an inhomogeneous
universe. 

In this paper, we investigate the relation between
distance and redshift in an inhomogeneous universe.
As an inhomogeneous model, we take a spherically symmetric
dust--shell model,
in which light rays travel from each dust shell
toward an observer at the center.
We calculate angular diameter distance--redshift 
($\dA$--$z$) relation
and compare it with that of a FL model.
Our main goal in this paper is to clarify the condition under which
the distance--redshift relation of 
a dust--shell universe 
behaves like that of a FL model.
We investigate the behavior of the averaged density around the 
observer at the center when we gradually extend the averaged region to
the outer shells, and
discuss the relation among the behavior of the averaged density,
the energy density of the FL model, and
the observed distance--redshift relation.

We note two interesting points of a dust--shell model.
First, its dynamics is exactly solved;
it is not necessary to assume the existence of 
homogeneous background in order to obtain the behavior 
of density fluctuations.
Secondly, it can treat a discrete mass distribution
where the linear perturbation theory is invalid, and 
can also treat a highly general-relativistic situation 
where the scale of inhomogeneities are comparable to 
the horizon scale. 

The organization of this paper is as follows.
In the next section, we present basic equations
for the dynamics of a dust--shell universe,
distance to the shells from the center,
and redshift of the shells measured by an observer at the center.
We give our results and discussion on $\dA$--$z$ relation 
and averaged density in section 3,
followed by concluding remarks in section 4.

We follow the sign convention of the Riemann tensor and
the metric tensor in \cite{rf:WALD} 
and adopt the unit of $c=1$.

\section{Formulation of Dust--shell universe}
\subsection{Equation of motion of dust--shell}
We put a number of spherically symmetric shells
whose centers are common at $r=0$ (see Figure 1). 
The innermost shell is called the first shell,  
the next one is called the second shell, and so on. 
The region enclosed by the $(i-1)$-th shell and $i$-th shell 
is called the $i$-th region. 
Each shell is infinitesimally thin, characterized by the 
surface stress-energy tensor which is given by
\begin{eqnarray}
  S^{ab} \equiv \lim_{\epsilon\rightarrow 0} 
\int_{-\epsilon}^{+\epsilon}
 T^{ab}\ dx \quad ,
\end{eqnarray}
where $x$ is a Gaussian coordinate ($x=0$ on the shell) in 
the direction normal to the shell.

Since each region between the shells is vacuum, 
the space-time is described by the Schwarzschild geometry 
by the Birkhoff's theorem. The line element in the $i$-th 
region is written in the form
\begin{eqnarray}
  ds_{i}^2 = - \left( 1 - {2Gm_{i}\over r}\right) dt^2 
  + \left( 1 -{2Gm_{i}\over r}\right)^{-1} dr^2 
  + r^2 d\Omega^2, \label{eq:st-line}
\end{eqnarray}
where the parameter $m_i$ will be 
referred to as a gravitational mass ($m_1 = 0$), and
$d\Omega$ is the line element of a unit sphere.

We first derive 
the expansion law of the $i$-th shell
following\cite{rf:MTW,rf:ISRAEL,rf:SATO,rf:MAEDA}.
Let $n^a$ be a unit space-like vector normal to the trajectory of the
shell, and 
define the projection operator $h_a^b \equiv \delta_a^b - n_a n^b$. 
From the projected Einstein equation 
\begin{eqnarray}
R_{ab} h^a_c h^b_d  =  8\pi G \left( T_{ab} - {1\over 2}g_{ab} T\right)
h^a_c h^b_d \quad ,
\end{eqnarray}
one obtains 
\begin{eqnarray}\label{proj}
  {\pounds}_n K_{cd} + {}^3 R_{cd} - K K_{cd} =
   8\pi G \left(T_{ab}h^a_c h^b_d - {1\over 2}h_{cd} T \right) \quad ,
\end{eqnarray}
where ${\pounds}_n$ is the Lie derivative along $n^a$
and ${}^3 R_{cd}$ is the 3-dimensional Ricci tensor on the 
timelike hypersurface generated by the motion of the shell.
The extrinsic curvature $K_{ab}$ is defined by 
$K_{ab} = - {1\over 2} h_a^c h_b^d
{\pounds}_n h_{cd}$, and $K = K_a^a$, $T= T_a^a$.
Integration of equation (\ref{proj}) 
over an infinitesimal range along $n^a$ yields
\begin{eqnarray}\label{K1}
  K_{ab}^+ - K_{ab}^- =  8\pi G
\left( S_{ab} - {1\over 2}h_{ab} S\right) \quad ,
\end{eqnarray}
where the suffix `$+$' denotes a quantity evaluated at the outside of
the shell and `$-$' at the inside.
Using equation (\ref{K1}) and the Gauss--Codazzi relation
$  2G_{ab} n^a n^b = - {}^3 R + K_{ab} K^{ab}- K^2 $,
one finds that the following relation holds for a dust--shell:
\begin{eqnarray}\label{K2}
  S^{ab} \left( K_{ab}^+ + K_{ab}^- \right) = 0 \quad .
\end{eqnarray}

Combining equations (\ref{K1}) and (\ref{K2})
and substituting the expression of the metric and the stress--energy
tensor,
we obtain the following equation for the circumferential radius $r_i$
(the ``expansion law'' of the dust shell):
\begin{eqnarray}\label{motion}
 \left({dr_{i}\over d\tau}\right)^2 = {2GM_+(i)\over r_i} 
+ \left\{ \left({M_-(i)\over m_{\rm s}(i)}\right)^2 -1\right\}
+ { G^2 m_{\rm s}^2(i)\over 4 r_i^2}\quad ,
\end{eqnarray}
where $M_{\pm}$ is defined by
\begin{eqnarray}
  M_+ (i) \equiv {m_i + m_{i+1}\over 2} \quad, \label{eq:gmass-def}
\end{eqnarray}
\begin{eqnarray}
  M_- (i) \equiv  m_{i+1}-m_i \quad ,\label{eq:dmass-def}
\end{eqnarray}
and $\tau$ is the proper time of the shell.
We have also introduced the ``baryonic'' mass $\ms (i)$ 
given by $m_{\rm s}(i) = 4 \pi s_i r_i^2 $
where $s_i$ is the surface density of the $i$-th shell;
$s_i = -S^a_a (i)$. 
It can be shown this baryonic mass is constant of motion 
by the conservation law, ${S_a^b}_{;b} = 0$ where
the semicolon denotes the 3-dimensional covariant derivative on the 
trajectory of the shell.

In this paper, we shall use a common proper time $\tau$ for 
all the shells. The relation between $\tau$ and 
the time coordinate, $t$, in Schwarzschild space-time is obtained as
follows. First, note that two Schwarzschild time coordinates
are assigned to  
each shell: the $i$-th shell has the time $t_{(-)i}$ measured 
in the $i$-th region and $t_{(+)i}$ measured in the $(i+1)$-th 
region. 
Then, from the normalization condition for the 4-velocity of a shell, 
we obtain
\begin{eqnarray}
 {dt_{(+)i}\over d\tau}
&=&\left({r_{i}\over r_{i}-2Gm_{i+1}}\right)
  \left[1-{2Gm_{i+1}\over r_{i}}
 +\left({dr_{i}\over d\tau}\right)^{2}\right]^{1\over2}, \label{eq:ex-time}\\
 {dt_{(-)i}\over d\tau}
&=&\left({r_{i}\over r_{i}-2Gm_{i}}\right)
  \left[1-{2Gm_{i}\over r_{i}}
 +\left({dr_{i}\over d\tau}\right)^{2}\right]^{1\over2}.\label{eq:in-time}
\end{eqnarray}
The procedure to determine the initial points of each time coordinate is
described later. 
%
%
\subsection{The solution and initial condition}
In order to specify a dust--shell universe, we have to fix
the parameters in equation (\ref{motion}) and the initial hypersurface.
We first rewrite equation (\ref{motion}) in the form
corresponding to the Hubble equation of FL models.
We denote the initial circumferential radius of the $i$-th shell 
by $x_{i}$, i.e., 
\begin{equation}
r_{i}=x_{i}~~~~~{\rm on~~initial~~hypersurface}.
\end{equation}
We define $\rho_i$ by
\begin{eqnarray}
 M_{+}(i) \equiv {4\over 3}\pi \rho_i {x_i}^3 \quad,
\label{eq:g-mass}
\end{eqnarray}
and $k_{i}$ by
\begin{eqnarray}\label{eq:defk}
\left({M_-(i)\over m_{\rm s}(i)}\right)^2  
\equiv 1-k_{i}x_{i}^{2} \quad .
\end{eqnarray}
Then the expansion law of the dust--shell is written as
\begin{eqnarray}\label{expa}
 \left( {1\over r_{i}}{dr_{i}\over d\tau}\right)^2 
 = {8\over 3} \pi G \rho_i
 \left( {x_i \over r_i}\right)^3  
 -k_{i}\left({x_{i}\over r_i}\right)^{2}
 + { G^2 m_{\rm s}^2(i)\over 4 r_i^4} \quad .
\end{eqnarray}
We see that the first term behaves like a non-relativistic matter term
in the Hubble equation of FL models, the second and the third
like a curvature and a radiation source term.
From this point of view, $\rho_i$ and $k_i$ play roles
of the ``energy density'' and the ``curvature'', respectively.
The radiation-like term might be regarded as the effect 
of the binding energy of the shell\cite{rf:SATO}. 
Further it is worthwhile to note that this radiation-like term 
is consistent with Futamase's result about the effect of the small-scale 
inhomogeneities on the global cosmic expansion\cite{rf:FUTAMASE}. 
Seeing this, one may expect that the inhomogeneities tend to make 
the Hubble parameter larger 
compared with a homogeneous universe
which has the same ``energy density'' of non-relativistic matter.
However, this radiation-like term 
does not necessarily imply the larger Hubble parameter.
In order to see the effect of this term
on the Hubble parameter, we need to investigate 
the distance--redshift relation 
by solving the null geodesic equations and compare 
the result in the inhomogeneous space-time and that 
of the FL model. Such an analysis will be performed 
in the following sections.

A dust--shell universe is specified
if we set the parameters contained in equation (\ref{expa}), i.e.,
$\rho_i$, $k_i$, $x_i$, and an initial hypersurface.
When we increase the number of the shells to infinity
with $\rho_i$ and $k_i$ being finite and independent of $i$
(we will mention this limit as ``large $N$ limit''),
the dust--shell universe approaches a FL model
if we take an appropriate initial hypersurface, as we will see in
section 3.
Then the parameters $\rho_i$ and $k_i$ agree with the ordinary
energy density and curvature in the Hubble equation.

We take 
\begin{eqnarray}
  k_i = 0 \label{eq:curvature}
\end{eqnarray}
for all $i$ in the remaining sections.
This means that the increase in the gravitational mass $m_i$ is 
equal to the baryonic mass of the shell.
In other words, the kinetic energy of the shell balances with the
potential energy, and hence the total energy becomes equal to the rest
mass. This is also the simplest case which approaches a flat FL model 
in the large $N$ limit.
Choice of the other parameters and the initial hypersurface
will be discussed in section 3.

For notational convenience, we shall 
introduce the following quantities:
\begin{equation}
\mu_{i}\equiv 2Gm_{i}, ~~~
 \nu_{i}\equiv {1\over2}G\ms (i)~~~{\rm and}~~~
 \sigma_{i}\equiv 2GM_{+}(i).
\end{equation}
From equations (\ref{eq:gmass-def}) and (\ref{eq:dmass-def}), 
the following relations hold
\begin{eqnarray}
\mu_{i}&=&\sigma_{i}-2\nu_{i}, \\
\mu_{i+1}&=&\sigma_{i}+2\nu_{i}.
\end{eqnarray}
Then the equation for the circumferential 
radius, $r_{i}$, of the $i$-th shell is written in the form
\begin{equation}
{dr_{i}\over d\tau}={1\over r_{i}}\sqrt{\sigma_{i}r_{i}+\nu_{i}^{2}},
\label{eq:initial}
\end{equation}
where we have assumed that each shell initially expands. 
From equations (\ref{eq:ex-time}) and (\ref{eq:in-time}), 
the equations for the Schwarzschild time 
coordinates, $t_{(\pm)i}$, are obtained as
\begin{equation}
{dt_{(\pm)i}\over d\tau}={r_{i}\mp\nu_{i}\over 
r_{i}-\sigma_{i}\mp2\nu_{i}}. \label{eq:time-eq}
\end{equation}
Thus, from equations (\ref{eq:initial}) and (\ref{eq:time-eq}), 
the equations for the relations between 
$t_{(\pm)i}$ and $r_{i}$ are given by
\begin{equation}
 {dt_{(\pm)i}\over dr_{i}}
={r_{i}(r_{i}\mp\nu_{i}) \over (r_{i}-\sigma_{i}\mp 2\nu_{i})
\sqrt{\sigma_{i}r_{i}+\nu_{i}^{2}}}. \label{eq:t-r-eq}
\end{equation}
The above equations can be integrated easily to give the
solution for $r_{i}>\sigma_{i}\pm2\nu_{i}$ in the form
\begin{equation}
t_{(\pm)i}(r_{i})=(\sigma_{i}\pm2\nu_{i})\ln
{\sqrt{\sigma_{i}r_{i}+\nu_{i}^{2}}
-(\sigma_{i}\pm\nu_{i})\over
\sqrt{\sigma_{i}r_{i}+\nu_{i}^{2}}
+(\sigma_{i}\pm\nu_{i})}
+T_{(\pm)i}(r_{i})+{\cal T}_{(\pm)i},\label{eq:time-sol}
\end{equation}
where
\begin{equation}
T_{(\pm)i}(r_{i})={2\over3\sigma_{i}^{2}}
\left(\sigma_{i}r_{i}+3\sigma_{i}^{2}\pm3\nu_{i}\sigma_{i}-2\nu_{i}^{2}\right)
\sqrt{\sigma_{i}r_{i}+\nu_{i}^{2}} \quad ,
\end{equation}
and ${\cal T}_{(\pm)i}$'s are integration constants. 

There is a coordinate singularity on the Killing horizon; 
$t_{(\pm)i}$ becomes infinite on $r_{i}=\sigma_{i}\pm2\nu_{i}$. 
For further calculation the null coordinate is convenient 
since we are interested in the null geodesics 
in this space-time.
Hence, we shall adopt the Kruscal null coordinates which has no 
coordinate singularity. 
Outside the horizon in the $i$-th region, $r>\mu_{i}$, 
the Kruscal null coordinate is given by
\begin{eqnarray}
U&\equiv&-2\sqrt{2\mu_i}(r-\mu_{i})^{1\over2}
  \exp\left(-{t-r\over2\mu_{i}}\right), \\
V&\equiv&+2\sqrt{2\mu_{i}}(r-\mu_{i})^{1\over2}
  \exp\left(+{t+r\over2\mu_{i}}\right),
\end{eqnarray}
where $U$ and $V$ correspond to the retarded time and
 the advanced time. Using these Kruscal coordinates, 
the line element in the $i$-th region is expressed as
\begin{equation}
ds_{i}^{2}=-{\mu_{i} \over 2r}
\exp\left(-{r \over \mu_{i}}\right)dUdV+r^{2}d\Omega^{2}.
\end{equation}

Similarly to the Schwarzschild time coordinate,
two pairs of Kruscal null coordinates, 
$U_{(\pm)i}$ and $V_{(\pm)i}$, are assigned to each shell.
Using equation (\ref{eq:time-sol}), 
we obtain the Kruscal null coordinates labeling the 
$i$-th shell in the form
\begin{eqnarray}
 U_{(\pm)i}(r_{i})
&=&-2\sqrt{2}\sqrt{1\pm{2\nu_{i}\over \sigma_{i}}}
 \left(\sqrt{\sigma_{i}r_{i}+\nu_{i}^{2}}+\sigma_{i}\pm\nu_{i}\right)
 \exp\left[{r_{i}-T_{(\pm)i}-{\cal T}_{(\pm)i}\over2(\sigma_{i}\pm2\nu_{i})}\right], \\
 V_{(\pm)i}(r_{i})
&=&+2\sqrt{2}\sqrt{1\pm{2\nu_{i}\over \sigma_{i}}}
 \left(\sqrt{\sigma_{i}r_{i}+\nu_{i}^{2}}-\sigma_{i}\mp\nu_{i}\right)
 \exp\left[{r_{i}+T_{(\pm)i}+{\cal T}_{(\pm)i}\over2(\sigma_{i}\pm2\nu_{i})}\right].
\end{eqnarray}
As expected, $U_{(\pm)i}$ and $V_{(\pm)i}$ are finite 
on $r_{i}=\sigma_{i}\pm2\nu_{i}$ and are well defined 
also for $r_{i}<\sigma_{i}\pm2\nu_{i}$.  
When $\sigma_{i}\pm2\nu_{i}$ is larger than $r_{i}$,
both $U_{(\pm)i}$ and $V_{(\pm)i}$ are negative. 
This means that the $i$-th shell with $r_{i}<\sigma_{i}\pm2\nu_{i}$ 
is located in the white hole part of the Schwarzschild 
space-time. 
This situation occurs for the shells beyond the horizon scale.

The determination of the integration constants ${\cal T}_{(\pm)i}$
corresponds to the choice of the initial hypersurface. 
The procedure to construct the initial hypersurface we adopt
is summarized as follows; First, we choose a unit spacelike vector
$\ell^a$  
which is directed outward in the ordinary sense with respect to 
$r$.
Taking this vector as a starting tangential vector,
we extend a spacelike geodesic curve until it reaches 
the second shell.
This spacelike geodesic curve defines the simultaneous hypersurface 
in the region between the first shell and the second shell.
Next we extend from this intersection towards the third shell
another spacelike geodesic 
which starts from another spacelike vector at the second shell.
This second spacelike geodesic 
generates a spacelike hypersurface in this region. 
Repeating this process, we complete the whole initial hypersurface.

From the above procedure, the integration constant of equation
(\ref{eq:time-sol}) is 
determined as follows. In 
the $i$-th region,  
the tangent vector of the spacelike geodesic is written as
\begin{equation}
\ell^{t}=E_{i}\left(1-{\mu_{i}\over r}\right)^{-1}~~~~~{\rm and}~~~~~
\ell^{r}=\sqrt{1+E_{i}^{2}-{\mu_{i}\over r}}, \label{eq:sg-comp}
\end{equation}
and the other components vanish, where $E_{i}$ is an integration 
constant associated with the geodesic equation and will be determined 
by the condition which we will see in the next section.
From equation (\ref{eq:sg-comp}), the equation for the 
trajectory of the spacelike geodesic in the $(t,r)$ plane 
is given by
\begin{equation}
{dt\over dr}={E_{i}r^{3\over2} \over 
(r-\mu_{i})\sqrt{(1+E_{i}^{2})r-\mu_{i}}} \quad .
\end{equation}
Integrating the above equation, we obtain
\begin{equation}
t=F_{i}(r)+D_{i},
\end{equation}
where 
\begin{eqnarray}
F_{i}(r)=
{E_i\sqrt{r}\sqrt{(1+E_i^2)r-{\mu_i}}\over 1+E_i^2}
+
{\mu_i} \ln\left({
\sqrt{(1+E_i^2)r-{\mu_i}}-E_i\sqrt{r}\over
\sqrt{(1+E_i^2)r-{\mu_i}}+E_i\sqrt{r}}\right)
\nonumber \\
+
{E_i(3+2E_i^2){\mu_i}
\ln(\sqrt{(1+E_i^2)r-{\mu_i}}+\sqrt{(1+E_i^2)r})\over 
  (1+E_i^2)^{3/2}}  \quad ,
\end{eqnarray}
and $D_{i}$ is an integration constant. 
Initially, we set $t_{(+)i}=t_{(-)i}$ and 
$t_{(\pm)1}=0$. 
Then, since $t_{(-)i-1}=F_{i-1}(x_{i-1})+D_{i-1}$ and 
$t_{(+)i-1}=F_{i}(x_{i-1})+D_{i}$, we find
\begin{equation}
D_{i}=D_{i-1}+F_{i-1}(x_{i-1})-F_{i}(x_{i-1}),
\end{equation}
and $D_{1}=-F_{1}(x_{1})$. 
From the above recurrent relation, we obtain
\begin{equation}
D_{i}=-F_{i}(x_{i-1})+\sum_{j=2}^{i-1}
\left[F_{j}(x_{j})-F_{j}(x_{j-1})\right].
\end{equation}
Using these relations, we obtain the integration constants, 
${\cal T}_{(\pm)i}$, for $i\geq2$ as
\begin{equation}
{\cal T}_{(\pm)i}=\sum_{j=2}^{i}
\left[F_{j}(x_{j})-F_{j}(x_{j-1})\right]
-(\sigma_{i}\pm2\nu_{i})\ln
{\sqrt{\sigma_{i}x_{i}+\nu_{i}^{2}}
-(\sigma_{i}\pm\nu_{i})\over
\sqrt{\sigma_{i}x_{i}+\nu_{i}^{2}}
+(\sigma_{i}\pm\nu_{i})}
-T_{(\pm)i}(x_{i}),
\end{equation}
and for $i=1$ as
\begin{equation}
{\cal T}_{(\pm)1}=-(\sigma_{1}\pm2\nu_{1})\ln
{\sqrt{\sigma_{1}x_{1}+\nu_{1}^{2}}
-(\sigma_{1}\pm\nu_{1})\over
\sqrt{\sigma_{1}x_{1}+\nu_{1}^{2}}
+(\sigma_{1}\pm\nu_{1})}
-T_{(\pm)1}(x_{1}) \quad .
\end{equation}

\subsection{Redshift and diameter distance}

We consider a light ray which is emitted
from each shell toward an observer at the center. 
The light ray goes along a future 
directed ingoing radial null geodesic, where 
``ingoing'' refers to the direction 
from a shell
toward shells labeled by a smaller number. 

An ingoing radial null geodesic is specified by
a constant $V$ in the Kruscal null coordinate. 
The circumferential radius of the $i$-th shell 
when it intersects the null geodesic is denoted by $R_{i}$.  
The outermost shell considered here is labeled by $M$. 
Then $R_{M}=x_{M}$ and hence in the $M$-th region, 
$V=V_{(-)M}(x_{M})$ is satisfied along the null geodesic. 
Thus, on the $(M-1)$-th shell, the following relation holds 
\begin{equation}
V_{(+)M-1}(R_{M-1})=V_{(-)M}(x_{M}).
\end{equation}
This equation determines $R_{M-1}$. We obtain 
the circumferential radii of all the shells at the intersection 
with the null geodesic by the same procedure, i.e., 
by solving the following recurrent relation:
\begin{equation}
V_{(+)i}(R_{i})=V_{(-)i+1}(R_{i+1}). \label{recurrent}
\end{equation}
We can determine $R_{i}$ from the given $R_{i+1}$
through this equation.

In order to derive the expression of redshift, we first write down the
components of the null geodesic tangent  
in the $i$-th region, $k^{\mu}(i)$, which is given in 
the Kruscal null coordinate as  
\begin{equation}
k^{U}(i)={2r\over \mu_{i}}
\exp\left({r\over\mu_{i}}\right)\omega_{i},
\end{equation}
and the other components vanish, 
where $\omega_{i}$ is an integration constant associated 
with the geodesic equation. 
We require that the observed frequency
of the photon at each shell is uniquely determined.
The observed frequency, $\omega_{\rm ob}(i)$, 
at the $i$-th shell is given by
\begin{eqnarray}
\omega_{\rm ob}(i)
&=&-k_{\mu}(i)u_{(+)}^{\mu}(i)
={1\over2}\omega_{i}{dV_{(+)i}\over d\tau} \quad, 
\label{eq:freq1}\\
\omega_{\rm ob}(i+1) &=&-k_{\mu}(i)u_{(-)}^{\mu}(i+1)
={1\over2}\omega_{i}{dV_{(-)i+1}\over d\tau},
\label{eq:freq2}
\end{eqnarray}
where
\begin{equation}
{dV_{(\pm)i}\over d\tau}= {\sqrt{2}
(\sqrt{\sigma_{i}R_{i}+\nu_{i}^{2}}\mp\nu_{i} )\over 
\sqrt{\sigma_{i}^2 \pm 2 \nu_{i}\sigma_{i}}}
\exp\left({R_{i}+T_{(\pm)i}(R_{i})+{\cal T}_{(\pm)i} 
\over 2(\sigma_{i}\pm 2 \nu_{i})}\right).
\end{equation}
Equations (\ref{eq:freq1}) and (\ref{eq:freq2}) gives the relation
between $\omega_{\rm ob}(i)$ and  
$\omega_{\rm ob}(i-1)$, for $i\geq2$, as 
\begin{equation}
\omega_{\rm ob}(i)=f(i)\omega_{\rm ob}(i-1) \quad ,
\end{equation}
where
\begin{equation}
f(i) \equiv {dV_{(-)i}/d\tau \over dV_{(+)i-1}/d\tau} \quad .
\end{equation}
For the first region, a direct calculation leads to 
\begin{equation}
\omega_{\rm ob}(1)={\omega_{\rm ob}(0)\over R_{1}}
\left(R_{1}+\nu_{1}+\sqrt{\sigma_{1}R_{1}
+\nu_{1}^{2}}\right)
\equiv f(1) \omega_{\rm ob}(0)
\end{equation}
where $\omega_{\rm ob}(0)$ is the frequency of 
the light ray observed by an observer rest at the origin $r=0$. 
Thus, using the above relations, we obtain the 
redshift of the light ray emitted from the $i$-th shell 
toward the observer rest at $r=0$ in the form
\begin{equation}
  1+z(i) = {\omega_{\rm ob}(i)\over \omega_{\rm ob}(0)}
=\prod_{j=1}^{i}f(j)\quad . \label{redshift}
\end{equation}

Our next task is to find the angular diameter--distance $\dA$ .
The definition of $\dA$ is
\begin{eqnarray}
  \dA \equiv {D\over \theta}
\end{eqnarray}
where $D$ is the physical length of the source perpendicular to the
line of sight, and $\theta$ is the observed angular size.
Since the space we are considering is spherically symmetric,
the diameter distance from the observer at the center
to the $i$-th shell
agrees with the circumferential radius $R_i$ 
when the null geodesic intersects it;
\begin{eqnarray} 
  \dA (i) = R_i \quad . \label{diameter}
\end{eqnarray}

We calculate $\dA$--$z$ relation in the dust--shell universe
using the relations
(\ref{recurrent}), (\ref{redshift}) and (\ref{diameter}).

\section{Results and Discussion}
\subsection{Setting the parameters of dust--shell model}
As mentioned in the previous section, the choice of 
the parameters and the initial hypersurface determines the 
behavior of a dust--shell universe.
Since we are interested in cases which have a FL limit,
we set the parameters so that they approach a FL model 
in the large $N$ limit.

For most cases described below, we set the mass distribution of shells as
\begin{eqnarray}
\rho_i &=& \rho_{\rm c}\ {\rm (independent\ of}\ i{\rm )},
\label{eq:density} 
\end{eqnarray}
at the initial slice.
This is not a unique but the simplest case which goes to a FL model
in the large $N$ limit.
Using $\rho_{\rm c}$, we define 
 $\Hsi$ and $r_{\rm H}$ by
\begin{eqnarray}\label{Hubbleconst}
  \Hsi^2 \equiv r_{\rm H}^{-2}\equiv{8\pi G \over 3}\rho_{\rm c} \quad .
\end{eqnarray}
In terms of FL models, $\Hsi$ and $r_{\rm H}$ may be regarded as
the ``Hubble constant'' and ``Hubble horizon radius'' .
However, we cannot say anything at this stage about
what relation they have with the Hubble constant and horizon scale of
FL models. 

For $x_i$, we put 
\begin{eqnarray}
   x_i(\tau_{\rm init}) = i\Delta x 
\end{eqnarray}
with a constant interval $\Delta x$
\begin{equation}
\Delta x \equiv {\rH \over N_{\rm H}}
\end{equation}
where $N_{\rm H}$ is some positive integer.
Here note that $x_{N_{\rm H}}=\rH$ and the following relation holds
\begin{equation}
{2GM_{+}({N_{\rm H}})\over x_{N_{\rm H}}}=1.
\end{equation}
Hence $\rH$ corresponds 
to the ``mean Schwarzschild radius'' of the $\NH$-th and $(\NH+1)$-th 
regions. 

Before we proceed, we estimate the magnitude of the radiation-like term in
equation (\ref{expa}). 
From equations (\ref{eq:g-mass}) and (\ref{eq:gmass-def}) with
$m_{1}=0$, we find 
\begin{eqnarray}
Gm_{2n-1}&=&{1\over \NH^{3}}(n-1)^{2}(4n-1)\rH, 
\label{eq:g-mass-odd} \\
Gm_{2n}&=&{1\over \NH^{3}}n^{2}(4n-3)\rH,
\label{eq:g-mass-even}
\end{eqnarray}
where $n$ is a positive integer. 
From equations (\ref{eq:curvature}) (\ref{eq:defk}) and (\ref{eq:dmass-def}),
we find $\ms (i)=m_{i+1}-m_{i}$ 
and obtain
\begin{eqnarray}
Gm_{\rm s}(2n-1)&=&{6n^{2}-6n+1\over \NH^{3}}\rH, \label{eq:c-mass-odd} \\
Gm_{\rm s}(2n)&=&{6n^{2}\over \NH^{3}}\rH. \label{eq:c-mass-even}
\end{eqnarray}
Thus, when we consider a large $N$ limit with 
fixing $x_{i}=\rH(i/\NH)$, the baryonic mass $Gm_{\rm s}(i)={\cal
  O}(\NH^{-1})$  
is regarded as a small quantity, compared with $GM_{+}(i)$ and 
 $Gm_{i}$. That is, the radiation-like term is of order $ \NH^{-2}$
of the first term and can be neglected when $\NH$ is large.

\subsection{Distance-redshift relation in ``orthogonal'' model}
As described in the previous section, the choice of $E_i$ corresponds
to choice of initial hypersurface. 
In a FL model, the simultaneous hypersurface is
orthogonal to the trajectory of matter.
Thus, we try choosing $E_i$ so that the vector $\ell^a$ is
orthogonal to the trajectory of each shell
(we will refer to this choice as ``model A'').
From the condition $\ell_{a} u^{a}_{(+)}(i-1) =0$ at $r=r_{i-1}$,
we obtain
\begin{eqnarray}
  E_i = {1\over r_{i-1}}\sqrt{\sigma_{i-1} r_{i-1}+\nu_{i-1}^2} \quad .
\end{eqnarray}

In Figure 2 we plot the angular diameter distance as a 
function of redshift of each shell for various $N_{\rm H}$ with a
common $\Hsi$ in the case of model A.
The total number of shells $\NT$ is taken to be $3\NH$.
We emit a photon toward the center from 
each shell so that every photon reaches at the center
simultaneously.
We have obtained a 
surprising result that all the curves are quite similar 
even in the case when we put only two shells within the initial horizon
radius; the deviation among the curves is at most about $10\%$.
On the other hand, it can be also seen that the slope at higher
redshifts becomes steeper as 
we decrease the number of shells.

We compare the $\dA$--$z$ relation in the
dust--shell universe obtained above with that of
a FL model.
We here adopt a spatially flat FL model,
since the ``curvature'' term in the expansion law of a dust--shell
(\ref{expa}) vanishes.
Moreover, when $\NH$ is large, 
the ``radiation'' term is negligible compared to the first term.
Thus, in this paper, we focus on the cases with large $\NH(\geq 10)$,
and compare them with a spatially flat FL model filled only with 
non-relativistic matter.
The cases with small $\NH(<10)$  will be mentioned later.

The Hubble equation in this flat FL model is
\begin{eqnarray}
  H^2 \equiv \left( {\dot{a} \over a}\right)^2 = {8\over 3} \pi G 
\rho_{{\tiny \rm FL}} 
\end{eqnarray}
with $\rho_{{\tiny \rm FL}} \propto a^{-3}$. 
The relation between the redshift and the scale factor
in FL models is $1+z = a_0 /a$ where subscript ``$0$'' denotes a value when
the observer receives the photon.
Thus we can write the present Hubble parameter $H_{0}$
in terms of the initial Hubble parameter $H_{\rm i}$ and 
redshift $z_{\rm i}$ as
\begin{eqnarray}\label{FRW}
  H_0^2 = \Hi^2 (1+z_{\rm i})^{-3} \quad .
\end{eqnarray}
The $\dA$--$z$ relation in the flat FL model 
is calculated once we fix $\Hi$ and $z_{\rm i}$,
since the relation is determined solely by $H_0$.

The dashed line in Figure 2 shows the $\dA$--$z$ relation in the flat
FL model with $H_{\rm i} = \Hsi$.
The redshift of the initial hypersurface is identified with the
redshift of the outermost shell for the case $\NT = 3\NH =150$
, i.e., $z_{\rm i} = z(i=150)$.
The deviation of the dust--shell universe from the flat FL model 
with $\NT = 150$ amounts to about $2\%$ around the maximum of the curve.
We have confirmed that the deviation from the FL model becomes small
as we increase $\NH$.
For smaller $\NH$, however, the redshift of the outermost shell becomes 
small, resulting a larger $H_0$ from the relation (\ref{FRW}).
We show in Figure 3 the $\dA$--$z$ relation in the dust--shell universe 
in the case of $\NH = 10$(solid line) 
and that in the flat FL model 
with $H_{\rm i} = \Hsi$ and $z_{\rm i} = z(i=30)$
(long dashed line).
In this case, the deviation amounts to about $10\%$.
Moreover, the difference lies not only in the normalization
of $H_0$, but also in the shape of the curve.
To see this, we also plotted a FL curve (short dashed line) with a 
Hubble parameter changed by $10\%$: $H_{\rm i} = 0.9\times \Hsi$.
Note that the change in the Hubble parameter of the FL model
only results in the change in the normalization of the curve.
We see that the slope at higher redshifts is steeper in the 
dust--shell universe than in the FL model.
It is also clear from Figure 2 that this tendency becomes strong as we
decrease $\NH$. 

Thus, we can say that for large $\NH$, the flat FL model approximates the
dust--shell universe quite well, but
as we decrease $N_H$, this fit becomes poorer.

In the next subsection we discuss the reason of 
this behavior by studying the behavior of averaged density,
and try to reduce the deviation from FL models
without increasing $\NH$.
\subsection{Behavior of Averaged density}
We usually regard a FL model as a large-scale ``average'' of a
locally inhomogeneous universe.
We will study the 
relation between the results obtained in the previous section and an
``averaged'' density 
of the dust--shell universe.
We consider an averaged density around the observer at the 
center. The averaged density is defined by
 dividing the mass contained within some radius 
by the 3-volume on the hypersurface up to that radius.
We will study the behavior of the averaged density 
when the radius to take the average is gradually increased,
and discuss its relation with the distance--redshift relation.

First let us consider the 3-volume on the initial hypersurface defined 
in the previous section.
From equation (\ref{eq:sg-comp}), 
the intrinsic metric of the initial hypersurface is given by
\begin{equation}
d\ell^{2}_{i}={r \over (1+E^{2}_{i})r-\mu_{i}}dr^{2}+r^{2}d\Omega^{2}.
\end{equation}
Using the above line element, we obtain the 
spatial volume, ${\rm Vol}(i)$, of the $i$-th region ($i>1$) 
on the initial slice in the form
\begin{eqnarray}
 {\rm Vol}(i)
&=&4\pi\int_{r_{i-1}}^{r_{i}}{r^{5\over2}\over
 \sqrt{(1+E_{i}^{2})r-\mu_{i}}}dr \nonumber \\
&=&4\pi
\left[\sqrt{(1+E_i^2)r-{\mu_i}}\left({r^{5/2}\over
  3(1+E_i^2)}+{5r^{3/2}{\mu_i}\over 12(1+E_i^2)^2} 
+{5\sqrt{r}{\mu_i}^2\over 8(1+E_i^2)^3}\right)\right]^{r_i}_{r_{i-1}}\\
&+&
{5\pi {\mu_i}^3 \over 2(1+E_i^2)^{7/2} }
\ln \left({
\sqrt{(1+E_i^2)r_i-{\mu_i}} + \sqrt{(1+E_i^2)r_i}
\over
\sqrt{(1+E_i^2)r_{i-1}-{\mu_i}} + \sqrt{(1+E_i^2)r_{i-1}}
}\right) \quad .
\end{eqnarray}
For $i=1$, ${\rm Vol}(1)$ is equal to $4\pi r_{1}^{3}/3$.

Using this volume, we define the averaged density $\bar{\rho}(i)$
as follows. 
We sum the baryonic masses up to the $(i-1)$-th shell and half of the
$i$-th shell, 
and divided the sum by the 3-volume inside the $i$-th shell;
\begin{equation}
  \bar{\rho}(i) \equiv
      \left\{m_{\rm s}(i)/2+\sum_{j\le i-1} m_{\rm s}(j)\right\}
      {\Big /}\sum_{k\le i}{\rm Vol}(k) \quad .
\end{equation}
It seems natural to take this sum of masses in averaging,
since the motion of the $i$-th shell 
is approximately determined by sum of the gravitational mass inside the
shell (which agrees 
with the sum of baryonic masses up to $(i-1)$-th shell) and half of its
baryonic mass;
the numerator in the above definition is the same with $M_+(i)$ which
appears in the first term in the expansion law (\ref{motion}).
We plotted in Figure 4 the averaged density $\bar{\rho}(i)$
for the model described in the previous section.
As expected, $\bar{\rho}(i)$ is almost constant near $\rho_{\rm c}$,
which may explain the reason the deviation of $\dA$-$z$ relation
in the dust--shell models from the FL curves is small.

However, the averaged density slightly becomes smaller at the outside region
than $\rho_{\rm c}$ defined by the expansion law. 
The reason is explained as follows.
The volume, $V$, between $r_i$ and $r_{i+1} \equiv r_i+\Delta r$ is
expanded in terms 
of $\Delta r$ as 
\begin{eqnarray}
  V = 4\pi \int_{r_i}^{{r_i}+\Delta r} \{l^r(r)\}^{-1} r^2 dr
  = 4\pi \int_{r_i}^{{r_i}+\Delta r} \{l^r(r_i)\}^{-1} r^2 dr
  - 4\pi \int_{r_i}^{{r_i}+\Delta r}
  l^r(r_i)'\{l^r(r_i)\}^{-2} (r-r_i)r^2 dr 
  + \cdot \cdot \cdot
\end{eqnarray}
where ${l^r}' = d l^r/dr$. 
Substituting the expression of $l^r$,
we can expand $V$ in terms of $1/{\NH}$ to find 
\begin{eqnarray}
  V = 4\pi r_{i}^2 \Delta r\ (1+ {1\over 2} {i\over {\NH}^2}
  + \cdot \cdot \cdot ) \quad .
\end{eqnarray}
We see that the volume becomes larger than that of a homogeneous model.
This effect is significant especially when $i>\NH$, i.e., beyond the  
horizon scale, which explains the behavior of the averaged density 
at large $i$ in Figure 4.
From this figure, one may think we can reduce the deviation 
from the FL model 
by adjusting the expansion law so that the averaged density 
is a constant value.
We plotted the $\dA$--$z$ relation in this case (model B)
and the corresponding FL curve for $\NH =10$ in Figure 5 (the 
solid line and the long dashed line).
We adjust $\rho_i$ in the expansion law iteratively
so that the relation 
\begin{eqnarray}
  \bar{\rho}(i)  = {3\over 8\pi} \Hsi^2
\end{eqnarray}
is satisfied.
As a result, $\rho_i$ is not homogeneous; it
increases as $i$ increases (Figure 6).
From Figure 5, we see that the difference between the dust--shell
universe and the FL model remains the same as Figure 3,
although the averaged density is indeed a constant.
\subsection{Cases with homogeneous averaged density field}
We can take an initial hypersurface in which both $\rho_i$ and
$\bar{\rho}(i)$ is  
homogeneous
by requiring the relation
\begin{eqnarray}
  {\rm Vol}(i) = {4\pi\over 3}(r_i^3 - r_{i-1}^3 ) \quad .
\end{eqnarray}
We plot the $\dA$--$z$ relation of this model (model C) in Figure 7.
Comparing with Figure 2, we see that the difference among the curves
is reduced.
Figure 8 illustrates the good agreement between the $\dA$--$z$ relations of
the dust--shell model and a FL model for the case $\NH = 10$. 

Here we check whether the assumption of constant
interval $\Delta x = 
\rH/\NH$ is essential or not. We 
try some patterns of the initial circumferential radius of the shells
$x_i$, for instance,
\begin{eqnarray}
  x_{i+1} = x_i + h \times a^{i}
\end{eqnarray}
where $h$ and $a$ are some positive constants and $a\neq 1$.
When $a>1 (<1)$, the distribution of shells becomes sparser (denser)
for larger $i$.
We have found that the results are unchanged.
We also try another pattern
\begin{eqnarray}
  x_{2n} &=& 2n \times \Delta x \quad ,\label{int1}\\
  x_{2n+1} &=& (2n+1) \times \Delta x + a \Delta x \label{int2}\quad ,
\end{eqnarray}
with $a\neq 0$.
We plotted the result when $\NH = 10$, $\NT = 30$, and $a = 0.05$
(model D) in Figure 9\footnote{There is a maximum
in $a$ according to $\NT$ when we impose the condition 
$\bar{\rho}(i) =\rho_i = {\rho_{\rm c}}$; if we increase $\NT$, 
the maximum value allowed for $a$ becomes small.
The value of $a$ adopted here
is about the maximum value for $\NT = 3\NH$.}.
We can see that the deviation from the FL model remains small.

We can take another interesting choice;
the expansion law is homogeneous (characterized by $\rho_{\rm c}$),
and the averaged density is also homogeneous,
but each quantity differs; $\rho_{\rm c} = a \bar{\rho}$
(model E).
This is realized by imposing a condition of the form
\begin{eqnarray}
  {\rm Vol}(i) = a {4\pi\over 3}(r_i^3 - r_{i-1}^3 ) \quad ,
\end{eqnarray}
with $a \neq 1$.
In Figure 10, we show the result in the case $\NH=10$ and $a=0.93$ 
(solid line).
The dashed lines are curves in a FL model.
The upper one uses the original ``Hubble
constant''(Eq.(\ref{Hubbleconst}), i.e., $\Hi =
\Hsi$, and the lower one uses the parameter changed by the 
same amount with the change in the volume, i.e., 
$\Hi^2 = \Hsi^2 \times a^{-1}$.
We see that the shape of the curve differs from a FL curve
(compare with Figure 8).
This indicates that the averaged density should agree with 
$\rho_{\rm c}$ which determines the expansion law, in order for the
$\dA$--$z$ relation 
to behave like that of a FL model.
It should be also noted that the FL curve with the changed Hubble parameter
is closer to the curve in dust--shell universe
than the other. This may indicate that the ``observed'' Hubble 
parameter is closer to the averaged density $\bar{\rho}$, rather than
$\rho_{\rm c}$ which determines the expansion law.
\subsection{Summary of models and Conclusions}
Here we summarize our results.
\begin{itemize}
\item Model A: The initial hypersurface is orthogonal to each trajectory
of shells and the expansion law is homogeneous; $\rho_i = \rho_{\rm c}$.
The $\dA$--$z$ relation of the dust--shell universe shows
deviation from the FL curve, especially when $\NH$ is small (Figs.2,3).
The averaged density $\bar{\rho}$ is not constant (Fig.4).

\item Model B: 
We adjust $\rho_i$ so that $\bar{\rho} = \rho_{\rm c}$
is satisfied (Fig.6) on the same initial hypersurface as Model A.
The deviation from the FL curve is not reduced (Fig.5).

\item Model C: We choose the initial hypersurface so that the relation
$\bar{\rho}(i)=\rho_i = \rho_{\rm c}$ is satisfied.
The $\dA$--$z$ relation of the dust--shell universe shows
good agreement with the FL curve (Figs.7,8).

\item Model D: The choice of the initial hypersurface is 
same with Model C, but the interval of the shells is not constant
(Eqs.(\ref{int1}),(\ref{int2})).
The deviation of $\dA$--$z$ relation of the dust--shell universe 
from the FL curve remains small (Fig.9).

\item Model E: The averaged density $\bar{\rho}(i)$ and the parameter
 $\rho_i$ are homogeneous but $\bar{\rho}(i)\neq \rho_i$.
The $\dA$--$z$ relation of the dust--shell universe shows
mild deviation from the FL curve (Fig.10).
\end{itemize}

From these results, we conclude that the $\dA$--$z$ relation
in a dust--shell universe looks like a flat FL universe,
when the expansion law resembles the flat FL model,
and the behavior of averaged density field is scale-independent
when we increase the scale of averaging,
and the averaged density agrees with $\rho_{\rm c}$.
This statement seems to be valid even in the cases with quite small
number of the shells.  
However, the situation is not so simple.
In small $\NH$ cases, the radiation-like term   
in the expansion law cannot be neglected.
One may expect that a FL model with radiation term
gives a better fitting to those cases, but
we have found this does not work.
This implies that we cannot tell 
the effect of homogeneities just by studying the expansion law.
We need more detailed study to this problem, which is left
for our future work.

We also note that in spatially flat cases
the gravitational mass and the baryonic mass coincide;
whether we use the baryonic mass or gravitational mass
in defining the averaged density, the result is the same.
In cases where those masses are different,
we have to be careful in determining the averaged density
when we try to construct a FL model which fits a dust--shell universe. 
In order to clarify which mass we should use to construct 
a fitting FL model,
we have to study non-zero $k_i$ cases, which will also be done
in our future paper.

\section{Summary}
We have studied the behavior of $\dA$--$z$ relation in a
spherically symmetric dust--shell universe
where the mass distribution is discrete.
We have compared the relation of dust--shell universe with that of FL models,
and discussed the relation with the behavior of averaged density.
We have seen that the $\dA$--$z$ relation observed at the center
agrees well with that of a flat FL model 
if the following conditions are satisfied:
(i) the expansion law of the circumferential radius of the shells
resembles the Hubble equation of a spatially flat FL model,  
(ii) the behavior of averaged density around the observer
at the center is scale-independent as we increase the scale on
which we take the average, and
(iii) the averaged density agrees with 
the energy density of the FL model.
We have also seen that the choice of initial hypersurface
relates the expansion law to the averaged density.

The effect of discreteness of mass distribution appears 
in the equation of motion of each dust--shell. This effect becomes
smaller as we increase the number density of shell.
We conclude the discreteness of matter distribution itself is of no
significance in this model
in discussing the observed quantities such as $\dA$ and $z$,
as long as the expansion law and the averaged density field is
homogeneous in the sense described above. 
This supports the averaging hypothesis that a universe
is described by a FL model if the universe is homogeneous 
when the density is averaged on some scale larger than the scale 
of the inhomogeneities. 

We need, however,
further discussion for the cases when the number of shells is 
extremely small, and when the curvature term 
does not vanish.
We also note that it will be interesting to study cases where 
the averaged density is inhomogeneous or non-radial null rays
travel, to see cosmological lensing effects.
These problems are left for our future work.

\section*{Acknowledgements}
We would like to thank H.~Sato for encouragement.
NS and TH are 
 supported by Research Fellowships of the Japan Society for 
the Promotion of Science for Young Scientists, No.3167
and No.9204.

%
\newpage
\begin{figure}
\epsfysize=8cm \epsfbox{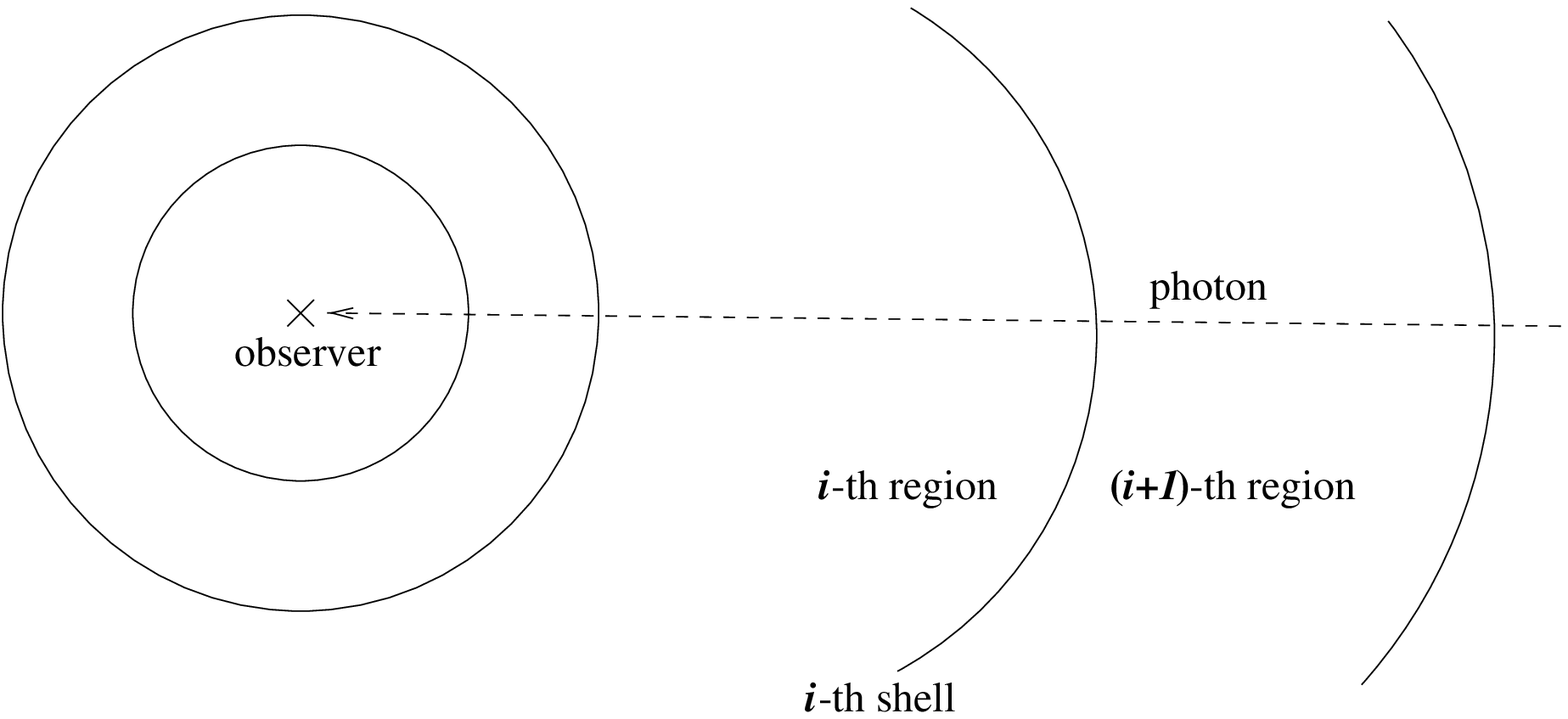}
\caption{Dust--shell universe.}
\end{figure}%
\begin{figure}[ht]
\epsfysize=8cm \epsfbox{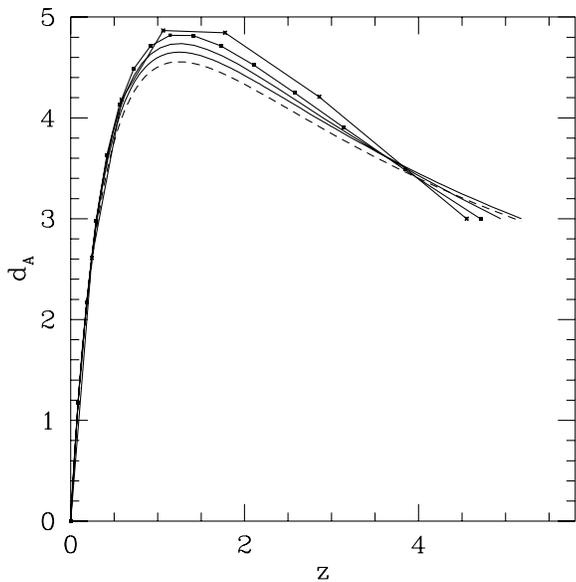}
\caption{Angular diameter distance--redshift relation in dust--shell universe
for model A.
Data points are connected by solid lines.
The definition of the models is summarized at the end of section 3.
The number of shells within the initial Hubble horizon $\NH$ 
is  $2,5,10,$ and $50$ from top to bottom around the maximum.
The total number of shells $\NT$ is taken to be $3\NH$.
We also mark the data by dots in the cases $\NH = 2$ and $5$.
We see that all the curves are quite similar;
the deviation among the curves is at most about $10\%$.
On the other hand, it can be also seen that the slope at higher
redshifts becomes steeper as 
we decrease the number of shells.
The dashed line shows the $\dA$--$z$ relation in the flat
FL model with $H_\protect{\rm i\protect} = \Hsi$.
The redshift of the initial hypersurface is identified with the
redshift of the outermost shell for the case $\NT =150$, i.e.,
$z_\protect{\rm i\protect} = z(i=150)$. 
The deviation between the flat FL model and the dust--shell universe
with $\NT = 150$ amounts to about $2\%$ around the maximum of the curve.
}
\end{figure}
\begin{figure}[t]
\epsfysize=8cm \epsfbox{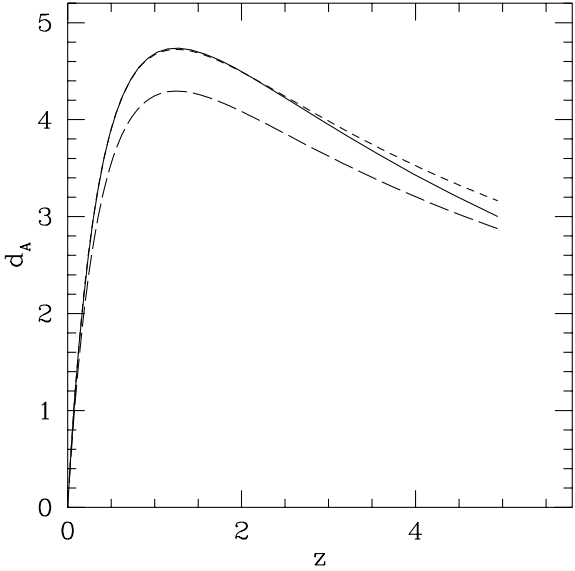}
\caption{Angular diameter distance--redshift relation in the
 dust--shell universe  
for model A in the case of $\NH = 10$(solid line) 
and that in the flat FL model 
with $H_\protect{\rm i\protect} = \Hsi$ and $z_\protect{\rm i\protect} =
z(i=30)$ 
(long dashed line).
We see that the deviation among them amounts to about $10\%$.
To see the difference in the shape of the curve, we also plotted a FL
curve with a  
Hubble parameter changed by $10\%$: $H_{\rm i} = \Hsi \times 0.9$ 
(short dashed line).
We see that the slope at higher redshifts is steeper in the 
dust--shell universe than in the FL model.
}
\end{figure}%
\begin{figure}[t]
\epsfysize=8cm \epsfbox{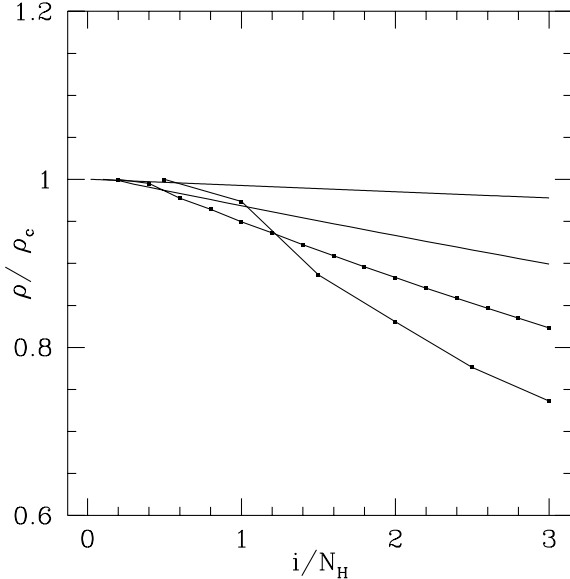}
\caption{Averaged density $\bar\protect{\rho\protect}$ normalized by
  $\rho_\protect{\rm c\protect}$ in the cases $\NH = 2,5,10$, and $50$ 
for model A. We connect the data by solid lines.
We also mark the data by dots in the cases $\NH = 2$ and $5$.
The curve which is smaller at large $i$
corresponds to the curve with smaller $\NH$.  
We see that the averaged density $\bar\protect{\rho\protect}(i)$ is almost
constant near 
$\rho_\protect{\rm c\protect}$, especially for small $i$, which may explain that the deviation
of $\dA$-$z$ relation 
in the dust--shell models from the FL curves is small.
One also notices that the averaged density becomes smaller at
the outside region than $\rho_\protect{\rm c\protect}$.
The interpretaion of this behavior is discussed in the text.
}
\end{figure}%
\begin{figure}[t]
\epsfysize=8cm \epsfbox{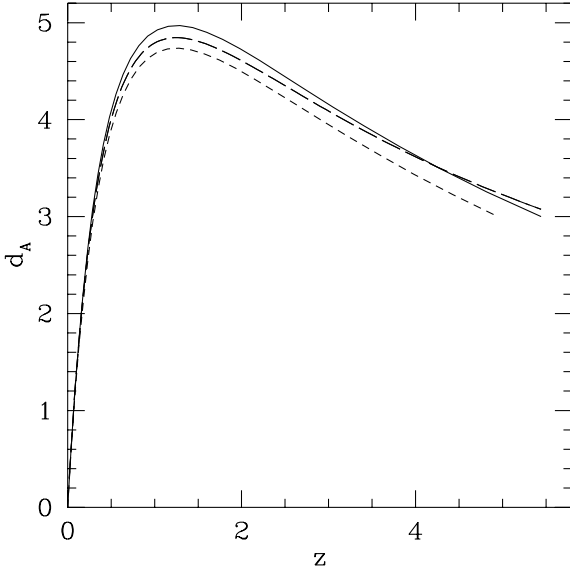}
\caption{Angular diameter distance--redshift relation in the
  dust--shell universe  
for model B in the case of $\NH = 10$(solid line) 
and that in the flat FL model 
with $H_\protect{\rm i\protect} = \Hsi$ and $z_\protect{\rm i\protect} =
z(i=30)$ 
(long dashed line).
For comparison, we also showed the curve in the dust--shell universe for
model A (short dashed line).
We see that the difference in the normalization 
between the dust--shell universe and the FL model
is reduced, but the difference in the shape of the curve
is not reduced compared with Figure 3 .
}
\end{figure}%
\begin{figure}[ht]
\epsfysize=8cm \epsfbox{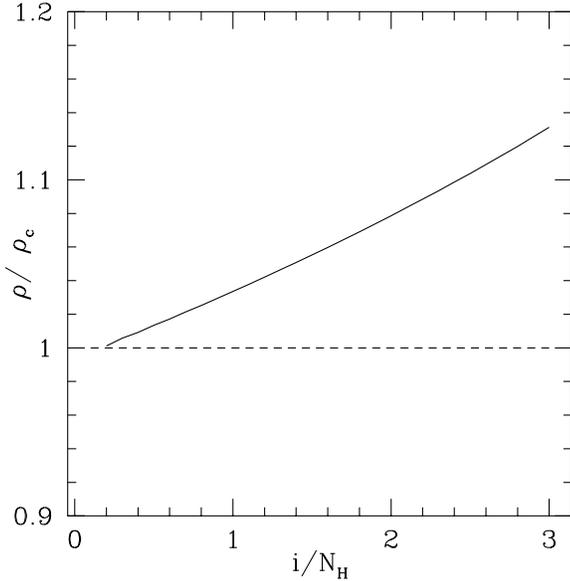}
\caption{
The averaged density $\bar\protect{\rho\protect}$ (dashed line) and 
the parameter $\rho_i$ (solid line) which determines the expansion law of
the shells of model B. 
We see that $\rho_i$ increases as $i$ increases,
while the averaged density $\bar\protect{\rho\protect}$ is indeed a constant.
}
\end{figure}
\begin{figure}[b]
\epsfysize=8cm \epsfbox{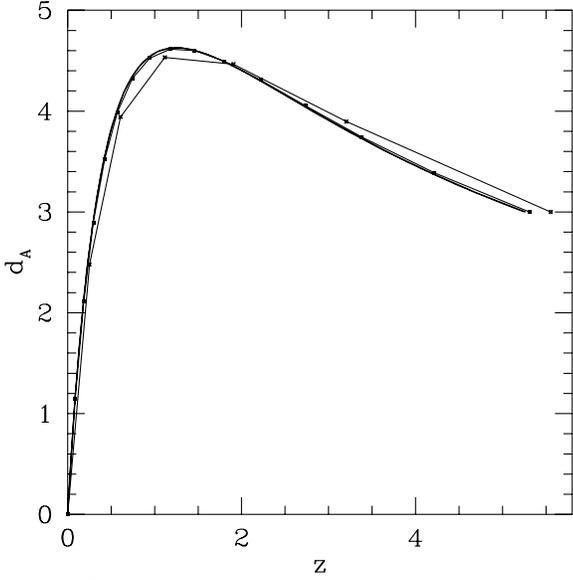}
\caption{Angular diameter distance--redshift relation in dust--shell universe
for model C.
The number of shells within the initial Hubble horizon $\NH$ 
is $2,5,10,$ and $50$.
The total number of shells $\NT$ is taken to be $3\NH$.
Data points are connect by solid lines.
We also mark the data by dots in the cases $\NH = 2$ and $5$.
Comparing with Figure 2, we see that the difference among the curves
is reduced so that 
it is hard to distingush the curves for $\NH = 5,10,$ and $50$.
}
\end{figure}%
\begin{figure}[t]
\epsfysize=8cm \epsfbox{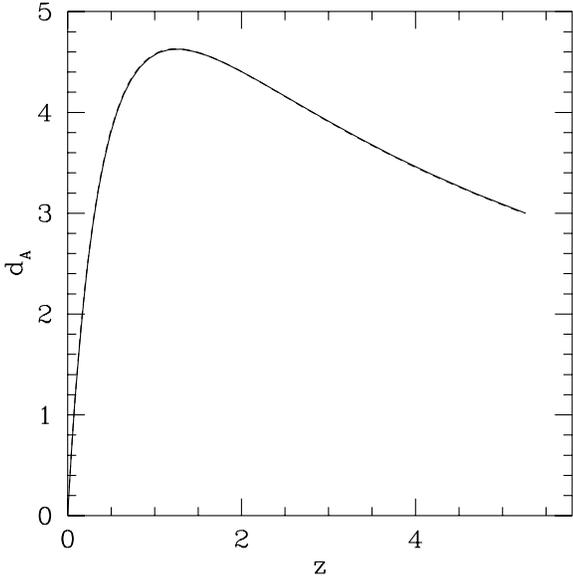}
\caption{Angular diameter distance--redshift relation in the
  dust--shell universe  
for model C in the case of $\NH = 10$(solid line) 
and that in the flat FL model 
with $H_\protect{\rm i\protect} = \Hsi$ and $z_\protect{\rm i\protect} =
z(i=30)$ 
(dashed line).
This illustrates the good agreement between the $\dA$--$z$ relations of
the dust--shell model and of a FL model for the case $\NH = 10$. 
}
\end{figure}
\begin{figure}[t]
\epsfysize=8cm \epsfbox{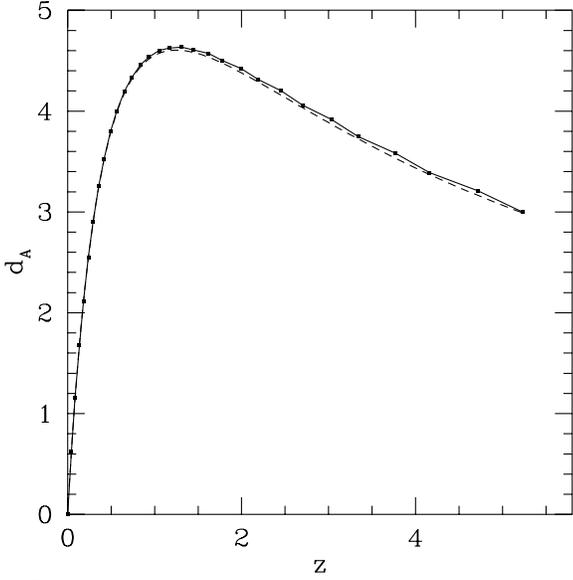}
\caption{Angular diameter distance--redshift relation in the
  dust--shell universe  
for model D in the case of $\NH = 10$(solid line and points) 
and that in the flat FL model 
with $H_\protect{\rm i\protect} = \Hsi$ and $z_\protect{\rm i\protect} =
z(i=30)$ 
(dashed line).
We can see that the deviation from the FL model remains small.
}

\end{figure}
\begin{figure}[b]
\epsfysize=8cm \epsfbox{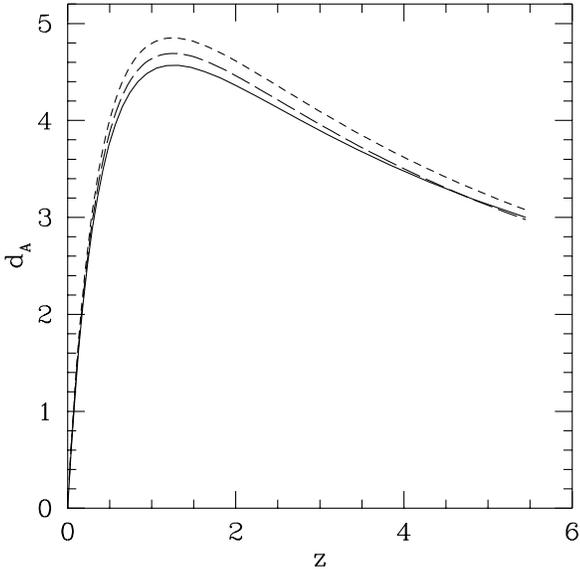}
\caption{Angular diameter distance--redshift relation in the
  dust--shell universe  
for model E in the case of $\NH = 10$(solid line)
and that in the flat FL model 
with $H_\protect{\rm i\protect} = \Hsi$ and $z_\protect{\rm i\protect} =
z(i=30)$ 
(short dashed line).
The long dashed line is the FL curve with the Hubble parameter changed
by the  same amount with the change in the volume, i.e., 
$\Hi^2 = \Hsi^2 \times a^{-1}$.
Comparing with Figure 8,
we see that the shape of the curve differs from a FL curve.
This indicates that the averaged density should agree with 
$\rho_\protect{\rm c\protect}$ which determines the expansion law, in order for the
$\dA$--$z$ relation to behave like that of a FL model.
Also note that the FL curve with the changed Hubble parameter
(long dashed line)
is closer to the curve in dust--shell universe
 than the other (short dashed line).
The interpretaion of this behavior is discussed in the text.
}
\end{figure}%
\end{document}